\documentstyle[aps,prd,floats]{revtex}
\input{psfig.tex}
\begin{document}
\draft

\title{An All-Sky Analysis of Polarization in the Microwave Background}
\author{Matias Zaldarriaga\footnote{matiasz@arcturus.mit.edu}}
\address{Department of Physics, MIT, Cambridge, Massachusetts~~02139}
\author{Uro\v s Seljak\footnote{useljak@cfa.harvard.edu}}
\address{Harvard-Smithsonian Center for Astrophysics,
60 Garden Street, Cambridge, Massachusetts~~02138}
\maketitle

\begin{abstract}
Using the formalism of spin-weighted functions we
present an all-sky analysis of polarization in 
the Cosmic Microwave Background (CMB). Linear polarization is a 
second-rank symmetric and traceless tensor, which 
can be decomposed 
on a sphere 
into spin $\pm 2$ spherical harmonics. These are 
the analog of the spherical harmonics
used in the temperature maps and obey the same completeness and 
orthogonality relations.
We show that there exist
two linear combinations of spin $\pm 2$ multipole moments
which have opposite parities and can be used to fully
characterize the statistical properties of polarization in the CMB.
Magnetic-type parity combination 
does not receieve contributions 
from scalar modes and does not cross-correlate
with either temperature or electric-type parity combination, so
there are four different power spectra that fully characterize statistical
properties of CMB.
We present their explicit expressions for scalar and tensor modes in 
the form of line of sight integral solution and
numerically evaluate them for a representative set of models.
These general solutions differ from the expressions obtained previously
in the small scale limit both for scalar and tensor modes.
A method to generate and analyze all sky maps of temperature and
polarization is given and
the optimal estimators for various power spectra 
and their corresponding variances are discussed. 

\end{abstract}


\def\edth{\;\raise1.0pt\hbox{$'$}\hskip-6pt\partial\;}
\def\baredth{\;\overline{\raise1.0pt\hbox{$'$}\hskip-6pt
\partial}\;}
\def\bi#1{\hbox{\boldmath{$#1$}}}
\def\gsim{\raise2.90pt\hbox{$\scriptstyle
>$} \hspace{-6.4pt}
\lower.5pt\hbox{$\scriptscriptstyle
\sim$}\; }
\def\lsim{\raise2.90pt\hbox{$\scriptstyle
<$} \hspace{-6pt}\lower.5pt\hbox{$\scriptscriptstyle\sim$}\; }

\section{Introduction}

The field of CMB anisotropies has become one of the main  
testing grounds for the theories of structure formation and early universe.
Since the first detection by COBE satellite \cite{smoot}
there have been several new detections on smaller 
angular scales (see \cite{reviews} for a recent review). 
There is hope that future
experiments such as  MAP \cite{map} and COBRAS/SAMBA \cite{cobra}
will accurately measure the anisotropies over the whole sky 
with a fraction of a degree angular resolution, which will help to 
determine several cosmological parameters with an unprecedented accuracy
\cite{jungman}.  
Not all of the cosmological 
parameters can be accurately determined by the CMB temperature
measurements. On large angular scales
cosmic variance (finite number of multipole moments on the sky)
limits our ability to extract useful
information from the observational data. If a certain parameter 
only shows its signature on large angular scales then the 
accuracy with which it can be determined is limited. For example, 
contribution from primordial gravity waves, if present, will only
be important on large angular scales. Because both scalar and 
tensor modes contribute to the temperature anisotropy one cannot 
accurately separate them  
if only a small number of independent realizations 
(multipoles) contain a significant contribution from tensor modes.
Similarly,
reionization tends to uniformly suppress the temperature 
anisotropies for all 
but the lowest multipole moments and is thus almost degenerate with 
the amplitude
\cite{jungman,bond94}.

It is clear from previous discussion that additional information
will be needed to constrain some of the cosmological 
parameters. While the epoch of reionization could in principle be 
determined through the high redshift observations, primordial gravity 
waves can only be detected at present from CMB observations. 
It has
been long recognized that there is additional information present 
in the CMB data in the form of linear polarization 
\cite{bond87,crittenden93,prev,coul,critt,zh95}.
Polarization
could be particularly useful for constraining the epoch and
degree of reionization because the amplitude is significantly increased
and has a characteristic signature \cite{zal}. 
Recently it was also shown that density perturbations (scalar modes)
do not contribute to polarization for a 
certain combination of Stokes 
parameters, in contrast with the primordial gravity waves
\cite{uros,letter,kks}, which can therefore in principle be detected even 
for very small amplitudes.
Polarization information which will potentially become
available 
with the next generation of experiments will thus provide
significant additional information that will help to
constrain the underlying cosmological model.

Previous work on polarization has been 
restricted to the small scale limit 
(e.g. \cite{crittenden93,prev,coul,uros,kosowsky96,polnarev}). 
The correlation
functions and corresponding power spectra 
were calculated for the Stokes $Q$ and $U$ 
parameters, which are defined with respect to a fixed coordinate 
system in the sky. While such a coordinate system is well defined over a
small patch in the sky, it becomes ambiguous once 
the whole sky is considered
because one cannot define a rotationally invariant orthogonal basis on a 
sphere. Note that this is not problematic if one is only considering 
cross-correlation function 
between polarization and temperature \cite{critt,coul}, 
where one can fix $Q$ or $U$ at a
given point and
average over temperature, which is rotationally invariant. However, if
one wants to analyze the auto-correlation function of polarization or
perform directly the power spectrum analysis on the data 
(which, as argued in \cite{uros}, is more efficient in terms of
extracting the signal from the data)
then a more general analysis of polarization is required.
A related problem is the calculation of rotationally invariant power spectrum.
Although it is relatively simple to
calculate $Q$ and $U$ in the coordinate system where the wavevector 
describing the perturbation is aligned with the $z$ axis, superposition
of the different modes becomes complicated because $Q$ and $U$ have to be
rotated to a common frame before the superposition can be done.
Only in the small scale limit can this rotation be simply expressed 
\cite{uros}, so
that the  
power spectra can be calculated. 
However, as argued above, this is not
the regime where polarization can make most significant impact 
in breaking the 
parameter degeneracies caused by cosmic variance.  A more general 
method that would allow to analyze polarization over the whole sky 
has been lacking so-far.

In this paper we present a complete all-sky analysis of polarization
and its corresponding power spectra. In section \S 2 we
expand polarization in the sky in
spin-weighted harmonics \cite{goldberg67,np}, which form a complete
and orthonormal system of tensor functions on the sphere. 
Recently, an alternative expansion in tensor harmonics 
has been presented \cite{kks}. Our approach differs both in the 
way we expand polarization on a sphere and in the way we solve
for the theoretical power spectra.
We use the line of sight integral solution 
of the photon Boltzmann equation \cite{sz}
to obtain the correct expressions for the polarization-polarization and 
temperature-polarization power spectra both for scalar (\S 3) and 
tensor (\S 4) modes. 
In contrast with previous work the expressions presented
here are valid for any angular scale and in \S 5 we 
show how they reduce to the corresponding small scale expressions.
In section \S 6 we discuss how to generate and analyze all-sky 
maps of polarization and what is the accuracy with which one can 
reconstruct the various power spectra when cosmic variance 
and noise are included. This is followed by discussion and 
conclusions in \S 7. 
For completeness we review in Appendix the basic properties of spin-weighted
functions. All the calculations in this paper are restricted to a flat 
geometry.

\section{Stokes parameters and spin-s spherical harmonics}

The CMB radiation field is characterized by a $2\times 2$ intensity 
tensor $I_{ij}$. The Stokes parameters $Q$ and $U$ are defined as
$Q=(I_{11}-I_{22})/4$ and $U=I_{12}/2$, while the temperature 
anisotropy is
given by $T=(I_{11}+I_{22})/4$. In principle the fourth  
Stokes parameter $V$ that describes circular polarization would also 
be needed, but in cosmology it can be ignored because it cannot
be generated through Thomson scattering.  
While the temperature is invariant
under a right handed rotation in the plane perpendicular to direction
$\hat{\bi{n}}$,
$Q$ and $U$ transform under rotation by an angle $\psi$ as
\begin{eqnarray}
Q^{\prime}&=&Q\cos 2\psi  + U\sin 2\psi  \nonumber \\  
U^{\prime}&=&-Q\sin 2\psi  + U\cos 2\psi 
\label{QUtrans} 
\end{eqnarray}
where ${\bf \hat{e_1}}^{\prime}=\cos \psi{\bf \hat{e_1}}+\sin\psi{\bf 
\hat{e_2}}$ 
and ${\hat{\bf e_2}}^{\prime}=-\sin \psi{\bf \hat{e_1}}+\cos\psi{\bf 
\hat{e_2}}$. 
This means we  can construct two quantities from the Stokes $Q$ 
and $U$ parameters that have 
a definite value of spin 
(see Appendix for a review of spin-weighted functions and their properties),  
\begin{equation}
(Q\pm iU)'(\hat{\bi{n}})=e^{\mp 2i\psi}(Q\pm iU)(\hat{\bi{n}}).
\end{equation}
We may therefore expand each of the quantities 
in the appropriate
spin-weighted basis 
\begin{eqnarray}
T(\hat{\bi{n}})&=&\sum_{lm} a_{T,lm} Y_{lm}(\hat{\bi{n}}) \nonumber \\
(Q+iU)(\hat{\bi{n}})&=&\sum_{lm} 
a_{2,lm}\;_2Y_{lm}(\hat{\bi{n}}) \nonumber \\
(Q-iU)(\hat{\bi{n}})&=&\sum_{lm}
a_{-2,lm}\;_{-2}Y_{lm}(\hat{\bi{n}}).
\label{Pexpansion}
\end{eqnarray}
$Q$ and $U$ are defined at a given direction $\bi{n}$
with respect to the spherical coordinate system $(\hat{{\bf e}}_\theta,
\hat{{\bf e}}_\phi)$.
Using the first equation in (\ref{propYs}) one can show that
the expansion coefficients for the polarization variables
satisfy $a_{-2,lm}^*=a_{2,l-m}$. For temperature the relation is
$a_{T,lm}^*=a_{T,l-m}$.

The main difficulty when computing the power spectrum of 
polarization in the
past originated in the fact that the Stokes parameters
are not invariant under
rotations in the plane perpendicular to $\hat{\bi{n}}$. While
$Q$ and $U$ are easily calculated in a 
coordinate system where the wavevector $\bi k$ is 
parallel to $\hat{\bi{z}}$,
the superposition 
of the different modes is complicated by the behaviour of $Q$ and $U$
under rotations (equation \ref{QUtrans}). For each wavevector
$\bi k$ and direction on the 
sky $\hat{\bi{n}}$ one has to rotate the $Q$ and $U$ parameters from the
$\bi{k}$ and $\hat{\bi{n}}$
dependent basis into a fixed basis on the sky. Only in the 
small scale limit is this process well defined, which is why this 
approximation has always been assumed in previous work 
\cite{crittenden93,prev,coul,uros,kosowsky96}. 
However, one can use the spin raising and lowering operators 
$\edth$ and $\baredth$ defined in Appendix
to obtain spin zero quantities. These  
have the advantage of being {\it rotationally invariant}
like the temperature and no ambiguities connected with the 
rotation of coordinate system arise. Acting twice with 
$\edth$, $\baredth$ on $Q\pm iU$ in equation (\ref{Pexpansion}) leads to 
\begin{eqnarray}
\baredth^2(Q+iU)(\hat{\bi{n}})&=&
\sum_{lm} \left[{(l+2)! \over (l-2)!}\right]^{1/2}
a_{2,lm}Y_{lm}(\hat{\bi{n}})
\nonumber \\
\edth^2(Q-iU)(\hat{\bi{n}})&=&\sum_{lm} \left[{(l+2)! \over (l-2)!}\right]^{1/2}
a_{-2,lm}Y_{lm}(\hat{\bi{n}}).
\end{eqnarray}

The expressions for the expansion coefficients are
\begin{eqnarray}
a_{T,lm}&=&\int d\Omega\; Y_{lm}^{*}(\hat{\bi{n}}) T(\hat{\bi{n}})
\nonumber  \\  
a_{2,lm}&=&\int d\Omega \;_2Y_{lm}^{*}(\hat{\bi{n}}) (Q+iU)(\hat{\bi{n}})
\nonumber  \\
&=&\left[{(l+2)! \over (l-2)!}\right]^{-1/2}
\int d\Omega\; Y_{lm}^{*}(\hat{\bi{n}}) 
\baredth^2 (Q+iU)(\hat{\bi{n}})
\nonumber \\  
a_{-2,lm}&=&\int d\Omega \;_{-2}Y_{lm}^{*}(\hat{\bi{n}}) (Q-iU)(\hat{\bi{n}}) 
\nonumber  \\  
&=&\left[{(l+2)! \over (l-2)!}\right]^{-1/2}
\int d\Omega\; Y_{lm}^{*}(\hat{\bi{n}})\edth^2 (Q-iU)(\hat{\bi{n}}).
\label{alm}
\end{eqnarray}

Instead of $a_{2,lm}$, $a_{-2,lm}$ it is convenient to introduce their
linear combinations
\cite{np}
\begin{eqnarray}
a_{E,lm}=-(a_{2,lm}+a_{-2,lm})/2 \nonumber \\ 
a_{B,lm}=i(a_{2,lm}-a_{-2,lm})/2.
\label{aeb}
\end{eqnarray}
These two combinations 
behave differently under parity transformation:  
while $E$ remains unchanged $B$ changes the sign \cite{np}, in 
analogy  
with electric and magnetic fields.
The sign convention in equation (\ref{aeb}) makes
these expressions consistent with those defined 
previously in the small scale limit \cite{uros}.

To characterize the statistics of the CMB perturbations
only four power spectra are needed, 
those for $T$, $E$, $B$ and the cross correlation between $T$ and $E$.
The cross correlation between $B$ and $E$ or $B$ and 
$T$ vanishes because 
$B$ has the opposite parity of $T$ and $E$.
We will 
show this explicitly for scalar and tensor modes in the following
sections.  
The power spectra are defined as the rotationally invariant quantities
\begin{eqnarray}
C_{Tl}&=&{1\over 2l+1}\sum_m \langle a_{T,lm}^{*} a_{T,lm}\rangle 
\nonumber \\
C_{El}&=&{1\over 2l+1}\sum_m \langle a_{E,lm}^{*} a_{E,lm}\rangle 
\nonumber \\
C_{Bl}&=&{1\over 2l+1}\sum_m \langle a_{B,lm}^{*} a_{B,lm}\rangle 
\nonumber \\
C_{Cl}&=&{1\over 2l+1}\sum_m \langle a_{T,lm}^{*}a_{E,lm}\rangle 
\label{Cls}
\end{eqnarray}
in terms of which,
\begin{eqnarray}
\langle a_{T,l^\prime m^\prime}^{*} a_{T,lm}\rangle&=&
C_{Tl} \delta_{l^\prime l} \delta_{m^\prime m} \nonumber \\
\langle a_{E,l^\prime m^\prime}^{*} a_{E,lm}\rangle&=&
C_{El} \delta_{l^\prime l} \delta_{m^\prime m} \nonumber \\
\langle a_{B,l^\prime m^\prime}^{*} a_{B,lm}\rangle&=&
C_{Bl} \delta_{l^\prime l} \delta_{m^\prime m} \nonumber \\
\langle a_{T,l^\prime m^\prime}^{*} a_{E,lm}\rangle&=&
C_{Cl} \delta_{l^\prime l} \delta_{m^\prime m} \nonumber \\
\langle a_{B,l^\prime m^\prime}^{*} a_{E,lm}\rangle&=&
\langle a_{B,l^\prime m^\prime}^{*} a_{T,lm}\rangle=
0. 
\label{stat}
\end{eqnarray}

For real space calculations it is useful to introduce  
two scalar quantities $\tilde{E}(\hat{\bi{n}})$ and $\tilde{B}(\hat{\bi{n}})$ 
defined as
\begin{eqnarray}
\tilde{E}(\hat{{\bi n}})&\equiv&
-{1\over 2}\left[\baredth^2(Q+iU)+\edth^2(Q-iU)\right]
\nonumber \\
	&=&\sum_{lm}\left[{(l+2)! \over (l-2)!}\right]^{1/2}
a_{E,lm}Y_{lm}(\hat{{\bi n}}) \nonumber \\  
\tilde{B}(\hat{\bi n})&\equiv&{i\over 2}
\left[\baredth^2(Q+iU)-\edth^2(Q-iU)\right]
\nonumber \\
	&=&\sum_{lm}\left[{(l+2)! \over (l-2)!}\right]^{1/2}
a_{B,lm}Y_{lm}(\hat{\bi n}) 
\label{EBexpansions} 
\end{eqnarray}
These variables have the advantage of being rotationally invariant
and easy to calculate in real space. These are not rotationally invariant
versions of
$Q$ and $U$, because $\edth^2$ and $\baredth^2$ are differential 
operators and are more closely related to the rotationally invariant
Laplacian of $Q$ and $U$. 
In $l$ space the two are simply related as
\begin{equation}
a_{(\tilde{E},\tilde{B}),lm}=\left[{(l+2)! \over (l-2)!}\right]^{1/2}
a_{(E,B),lm}.
\label{eblm}
\end{equation}

\section{Power Spectrum of Scalar Modes}

The usual starting point for solving the radiation transfer
is the Boltzmann equation. We will expand
the perturbations in Fourier modes characterized by wavevector $\bi{k}$.
For a given Fourier mode we can work in the 
coordinate system where 
$\bi{k} \parallel \hat{\bi{z}}$ and 
$(\hat{{\bf e}}_1,\hat{{\bf e}}_2)=(\hat{{\bf e}}_\theta,
\hat{{\bf e}}_\phi)$. 
For each plane wave the scattering can be described as the transport through
a plane
parallel medium \cite{chandra,kaiser}. 
Because of azimuthal symmetry only $Q$ 
Stokes parameter is generated in this frame and its amplitude
only depends on the 
angle between the photon direction and wavevector,
$\mu=\hat{\bi{n}}\cdot\hat{\bi{k}}$.
The Stokes parameters for this mode are
$Q=\Delta_P^{(S)}(\tau,k,\mu)$ and $U=0$, 
where the superscript $S$ denotes 
scalar modes,
while the temperature anisotropy is denoted with
$\Delta_T^{(S)}(\tau,k,\mu)$. The Boltzmann equation 
can be written in the synchronous gauge as \cite{bond87,mabert}
\begin{eqnarray} 
\dot\Delta_T^{(S)} +ik\mu \Delta_T^{(S)} 
&=&-{1\over 6}\dot h-{1\over 6}(\dot h+6\dot\eta)
P_2(\mu) +\dot\kappa\left[-\Delta_T^{(S)} +
\Delta_{T0}^{(S)} +i\mu v_b +{1\over 2}P_2(\mu)\Pi
\right] \nonumber \\   
\dot\Delta_P^{(S)} +ik\mu \Delta_P^{(S)} &=& \dot\kappa \left[
-\Delta_P^{(S)} +
{1\over 2} [1-P_2(\mu)] \Pi\right] \nonumber \\
\Pi&=&\Delta_{T2}^{(S)}
+\Delta_{P2}^{(S)}+
\Delta_{P0}^{(S)}.
\label{Boltzmann}
\end{eqnarray}
Here the derivatives are taken with respect to the conformal time $\tau$. 
The differential optical depth for Thomson scattering is denoted as 
$\dot{\kappa}=an_ex_e\sigma_T$, where $a(\tau)$ 
is the expansion factor normalized
to unity today, $n_e$ is the electron density, $x_e$ is the ionization 
fraction and $\sigma_T$ is the Thomson cross section. The total optical 
depth at time $\tau$ is obtained by integrating $\dot{\kappa}$,
$\kappa(\tau)=\int_\tau^{\tau_0}\dot{\kappa}(\tau) d\tau$.
The sources in these equations involve
the multipole moments of temperature and polarization, which 
are defined as $ \Delta(k,\mu)=\sum_l(2l+1)(-i)^{l}\Delta_l(k)P_l(\mu)$, 
where $P_l(\mu)$ is the Legendre polynomial of order $l$.
Temperature anisotropies have additional sources
in metric perturbations $h$ and $\eta$
and in baryon velocity term $v_b$.

To obtain the complete solution we need to evolve the anisotropies 
until the present epoch and integrate over all
the Fourier modes,
\begin{eqnarray}
T^{(S)}(\hat{\bi{n}})&=&\int d^3 \bi{k} \xi(\bi{k})\Delta_T^{(S)}(\tau=\tau_0,k,\mu) \nonumber \\
(Q^{(S)}+iU^{(S)})(\hat{\bi{n}})
&=&\int d^3 \bi{k} \xi(\bi{k})e^{-2i\phi_{k,n}}\Delta_P^{(S)}
(\tau=\tau_0,k,\mu) \nonumber \\
(Q^{(S)}-iU^{(S)})(\hat{\bi{n}})
&=&\int d^3 \bi{k} \xi(\bi{k})e^{2i\phi_{k,n}}\Delta_P^{(S)}
(\tau=\tau_0,k,\mu), 
\end{eqnarray}
where
$\phi_{k,n}$ is the angle needed to rotate 
the $\bi{k}$ and $\hat{\bi{n}}$ dependent basis to a
fixed frame in the sky. This rotation 
was a source of complications in previous
attempts to characterize the CMB polarization. We will avoid it in what 
follows by working with the rotationally invariant quantities.
We introduced $\xi(\bi{k})$, which 
is a random variable used to characterize the initial 
amplitude of the mode. It has the following statistical property
\begin{equation}
\langle \xi^{*}(\bi{k_1})\xi(\bi{k_2})
\rangle=
P_\phi(k)\delta(\bi{k_1}- \bi{k_2}),
\end{equation}
where $P_\phi(k)$ is the initial power spectrum. 

To obtain the power spectrum we 
integrate the Boltzmann equation (\ref{Boltzmann}) 
along the line of sight \cite{sz} 
\begin{eqnarray}
\Delta_T^{(S)}(\tau_0,k,\mu) &=& 
\int_0^{\tau_0} d\tau e^{ix \mu} S_T^{(S)}(k,\tau) \nonumber \\ 
\Delta_P^{(S)}(\tau_0,k,\mu) &=& {3 \over 4}(1-\mu^2)\int_0^{\tau_0} d\tau 
e^{ix \mu}g(\tau)\Pi(k,\tau) \nonumber \\
S_T^{(S)}(k,\tau)&=&g\left(\Delta_{T,0}+2 \dot{\alpha}
+{\dot{v_b} \over k}+{\Pi \over 4 }
+{3\ddot{\Pi}\over 4k^2 }\right)\nonumber \\
&+& e^{-\kappa}(\dot{\eta}+\ddot{\alpha})
+\dot{g}\left(\alpha+{v_b \over k}+{3\dot{\Pi}\over 4k^2 }\right)
+{3 \ddot{g}\Pi \over
4k^2} \nonumber \\
\Pi&=&\Delta_{T2}^{(S)}
+\Delta_{P2}^{(S)}+
\Delta_{P0}^{(S)},
\label{integsolsc}
\end{eqnarray}
where $x=k (\tau_0 - \tau)$ 
and $\alpha=(\dot h + 6 \dot \eta)/2k^2$.
We have introduced the visibility function $g(\tau)=\dot{\kappa}
{\rm exp}(-\kappa)$. Its peak  
defines the epoch of recombination, which gives the  
dominant contribution to the CMB anisotropies.

Because in the $\bi{k}\parallel \hat{\bi{z}}$  coordinate 
frame $U=0$ and $Q$ is only
a function of $\mu$ it follows from equation
(\ref{operators1}) that 
$\baredth^2(Q+iU)=\edth^2(Q-iU)$, so that ${}_2a_{lm}={}_{-2}a_{lm}$. 
Scalar modes thus contribute only to the $E$ combination and 
$B$ vanishes identically.  
Acting with the spin raising operator 
twice
on the integral solution for $\Delta_P^{(S)}$ (equation 
\ref{integsolsc}) leads to the following 
expressions for the scalar polarization $\tilde{E}$
\begin{eqnarray}
\Delta_{\tilde{E}}^{(S)}(\tau_0,k,\mu)&=&-{3 \over 4}
\int_0^{\tau_0} d\tau g(\tau)\Pi(\tau,k) 
\; \partial^2_{\mu} \left[(1-\mu^2)^2 e^{ix\mu}
\right] \nonumber \\
&=&{3 \over 4}\int_0^{\tau_0} d\tau g(\tau)\Pi(\tau,k)\; (1+\partial_x^2)^2 
\left(x^2 e^{ix\mu} \right). 
\label{tilEs}
\end{eqnarray} 

The power spectra defined in equation 
(\ref{Cls}) are rotationally invariant quantities
so they can be calculated in the frame 
where $\bi{k} \parallel \hat{\bi{z}}$
for each Fourier mode and then integrated over all the modes, 
as different modes are
statistically independent. The present day
amplitude for each mode depends both  
on its evolution and on its  
initial amplitude.
For temperature anisotropy $T$ it is given by \cite{sz}
\begin{eqnarray} 
C_{Tl}^{(S)}&=&
{1 \over 2l+1} \int d^3\bi{k} P_\phi(k)
\sum_m \left|\int d\Omega Y^*_{lm}(\hat{\bi{n}})
\int_0^{\tau_0} d\tau
S^{(S)}_T(k,\tau)
\; e^{ix\mu}\right|^2
\nonumber \\
&=&(4\pi)^2\int k^2dkP_\phi(k)\left[
\int_0^{\tau_0} d\tau
S^{(S)}_T(k,\tau)j_l(x) \right]^2 
\end{eqnarray}
where $j_l(x)$ is the spherical Bessel function of order $l$ and we 
used that in the $\bi{k} \parallel \hat{\bi{z}}$ frame
$\int d\Omega Y^*_{lm}(\hat{\bi{n}})\; e^{ix\mu}= 
\sqrt{4\pi(2l+1)}i^l j_l(x) \delta_{m0}$. 
For the spectrum of $E$ polarization the calculation is 
similar. Equation (\ref{tilEs}) is used to compute
the power spectrum of $\tilde{E}$ which combined with
equation (\ref{eblm}) gives
\begin{eqnarray} 
C_{El}^{(S)}&=&
{1 \over 2l+1} {(l-2)! \over (l+2)!}
\int d^3\bi{k} P_\phi(k)\sum_m \left|{3 \over 4}\int_0^{\tau_0} d\Omega Y^*_{lm}(\hat{\bi{n}})
\int_0^{\tau_0} d\tau
g(\tau)\Pi(k,\tau)
\; ([1+\partial_x^2]^2 (x^2e^{ix\mu})\right|^2
\nonumber \\
&=&
(4\pi)^2{(l-2)! \over (l+2)!}
\int k^2dk P_\phi(k)\left(
{3 \over 4}\int_0^{\tau_0} d\tau
g(\tau)\Pi(\tau,k)
\; ([1+\partial_x^2]^2 [x^2j_l(x)]\right)^2
\nonumber \\
&=&(4\pi)^2{(l+2)! \over (l-2)!}\int k^2dkP_\phi(k)\left[
{3 \over 4}\int_0^{\tau_0} d\tau
g(\tau)\Pi(\tau,k){j_l(x) \over x^2}\right]^2.
\end{eqnarray}
To obtain
the last expression we used the differential equation satisfied
by the spherical Bessel functions, $j_l^{\prime \prime}+2j_l^{\prime}/x+
[1-l(l+1)/x^2]j_l=0$.
If we introduce
\begin{eqnarray}
\Delta^{(S)}_{Tl}(k)&=&\int_0^{\tau_0} d\tau 
S^{(S)}_{T}(k,\tau) j_l(x) \nonumber \\  
\Delta^{(S)}_{El}(k)&=&\sqrt{(l+2)! \over (l-2)!}\int_0^{\tau_0} d\tau 
S^{(S)}_{E}(k,\tau) j_l(x) \nonumber \\  
S^{(S)}_E(k\tau)&=& 
{3g(\tau)\Pi(\tau,k) \over 4 x^2},
\label{es}
\end{eqnarray}
then the power spectra for $T$ and $E$ and their  cross-correlation 
are simply given by
\begin{eqnarray} 
C_{T,El}^{(S)}&=&
(4\pi)^2\int k^2dkP_\phi(k)\Big[\Delta^{(S)}_{T,El}(k)\Big]^2
\nonumber \\
C_{Cl}^{(S)}&=&
(4\pi)^2\int k^2dkP_\phi(k)\Delta^{(S)}_{Tl}(k)
\Delta^{(S)}_{El}(k).
\label{esc}
\end{eqnarray}
Equations (\ref{es}) and (\ref{esc}) are the main results of this section.

\section{Power spectrum of tensor modes}

The method of analysis used in previous section for scalar polarization 
can be used for tensor modes as well.
The situation is somewhat more complicated here because  
for each Fourier mode 
gravity waves have two independent polarizations usually 
denoted with  $+$ and $\times$. For our purposes it is convenient to
rotate this combination and work with the following two linear
combinations,
\begin{eqnarray}
\xi^1 &=&(\xi^+ - i \xi^\times)/ \sqrt{2} \nonumber \\ 
\xi^2 &=&(\xi^+ + i \xi^\times)/ \sqrt{2} 
\end{eqnarray}
where $\xi$'s are independent random variables
used to characterize the statistics of the gravity
waves. These variables have the following statistical
properties
\begin{equation}
\langle \xi^{1*}(\bi{k_1})\xi^1(\bi{k_2})
\rangle=\langle \xi^{2*}(\bi{k_1})\xi^2(\bi{k_2})
\rangle=
{P_h(k)\over 2}\delta(\bi{k_1}- \bi{k_2}),
\; \langle \xi^{1*}(\bi{k_1})\xi^2(\bi{k_2})
\rangle=0
\label{statxi}
\end{equation}
where $P_h(k)$ is the primordial power spectrum of 
the gravity waves.

In the coordinate
frame where $\hat{\bi{k}} \parallel \hat{\bi{z}}$ and 
$({\bf e}_1,{\bf e}_2)=({\bf e}_\theta,{\bf e}_\phi)$
tensor perturbations can be decomposed as 
\cite{kosowsky96,polnarev},
\begin{eqnarray}
\Delta_T^{(T)}(\tau,\hat{\bi{n}},\bi{k}) &=& \left[(1-\mu^2)e^{2i\phi}\xi^1(\bi{k})
+ (1-\mu^2)e^{-2i\phi}\xi^2(\bi{k})\right]
\tilde{\Delta}_T^{(T)}(\tau,\mu,k) \nonumber \\ 
(\Delta_Q^{(T)}+i\Delta_U^{(T)})
(\tau,\hat{\bi{n}},\bi{k}) &=& \left[(1-\mu)^2 e^{2i\phi}\xi^1(\bi{k})
+ (1+\mu)^2e^{-2i\phi}\xi^2(\bi{k})\right]
\tilde{\Delta}_P^{(T)}(\tau,\mu,k)\nonumber \\ 
(\Delta_Q^{(T)}-i\Delta_U^{(T)})(\tau,\hat{\bi{n}},\bi{k}) 
&=& \left[(1+\mu)^2 e^{2i\phi}\xi^1(\bi{k})
+ (1-\mu)^2e^{-2i\phi}\xi^2(\bi{k})\right]
\tilde{\Delta}_P^{(T)}(\tau,\mu,k),
\label{deconten}
\end{eqnarray}
where 
$\tilde{\Delta}_T^{(T)}$ 
and $\tilde{\Delta}_P^{(T)}$ are the variables introduced by
Polnarev to describe the temperature and polarization perturbations 
generated by gravity waves.
They satisfy the
following Boltzmann equation \cite{crittenden93,polnarev}
\begin{eqnarray} 
&\dot{\tilde{\Delta}}_T^{(T)}& +ik\mu \tilde{\Delta}_T^{(T)} 
=-\dot h
-\dot\kappa[\tilde{\Delta}_T^{(T)}-\Psi
] \nonumber \\   
&\dot{\tilde{\Delta}}_P^{(T)}& 
+ik\mu \tilde{\Delta}_P^{(T)} = -\dot\kappa [\tilde{\Delta}_P^{(T)} +
\Psi ] \nonumber \\
&\Psi & \equiv  \Biggl\lbrack
{1\over10}\tilde{\Delta}_{T0}^{(T)}
+{1\over 7}
\tilde {\Delta}_{T2}^{(T)}+ {3\over70}
\tilde{\Delta}_{T4}^{(T)}
 -{3\over 5}\tilde{\Delta}_{P0}^{(T)}
+{6\over 7}\tilde{\Delta}_{P2}^{(T)}
-{3\over 70}
\tilde{\Delta}_{P4}^{(T)} \Biggr\rbrack.
\label{BoltzmannT}
\end{eqnarray}
Just like in the scalar case these equations can be integrated along the
line of sight to give
\begin{eqnarray}
\Delta_T^{(T)}(\tau_0,\hat{\bi n},{\bi k}) &=& 
\left[(1-\mu^2)e^{2i\phi}\xi^1({\bi k})
+ (1-\mu^2)e^{-2i\phi}\xi^2({\bi k})\right]
\int_0^{\tau_0} d\tau e^{ix \mu} S_T^{(T)}(k,\tau) \nonumber \\ 
(\Delta_Q^{(T)}+i\Delta_U^{(T)})
(\tau_0,\hat{\bi n},{\bi k}) &=& \left[(1-\mu)^2 e^{2i\phi}\xi^1({\bi k})
+ (1+\mu)^2e^{-2i\phi}\xi^2({\bi k})\right]
\int_0^{\tau_0} d\tau 
e^{ix \mu} S_P^{(T)}(k,\tau) \nonumber \\ 
(\Delta_Q^{(T)}-i\Delta_U^{(T)})
(\tau_0,\hat{\bi n},{\bi k}) &=& \left[(1+\mu)^2 e^{2i\phi}\xi^1({\bi k})
+ (1-\mu)^2e^{-2i\phi}\xi^2({\bi k})\right]
\int_0^{\tau_0} d\tau 
e^{ix \mu} S_P^{(T)}(k,\tau) 
\label{integsolten}
\end{eqnarray}
where  
\begin{eqnarray}
S_T^{(T)}(k,\tau) &=& -\dot he^{-\kappa}+g\Psi \nonumber \\
S_P^{(T)}(k,\tau) &=& -g\Psi .
\label{sourten}
\end{eqnarray}

Acting twice with the spin raising and lowering operators on the 
terms with $\xi^1$ gives
\begin{eqnarray}
\baredth^2(\Delta_Q^{(T)}+i\Delta_Q^{(T)})(\tau_0,\hat{\bi{n}},\bi{k})&=&
\xi^1(\bi{k})e^{2i\phi}\int_0^{\tau_0} d\tau
S_P^{(T)}(k,\tau)\left(-\partial \mu + {2 \over 1-\mu^2}\right)^2 \left[
(1-\mu^2) (1-\mu)^2 e^{ix\mu}\right] \nonumber \\
	&=&\xi^1(\bi{k})e^{2i\phi}\int_0^{\tau_0} d\tau
S_P^{(T)}(k,\tau)[-{\hat{\cal E}}(x)-i{\hat{\cal B}}(x)]\left[
(1-\mu^2) e^{ix\mu}\right] \nonumber \\
\edth^2(\Delta_Q^{(T)}-i\Delta_Q^{(T)})(\tau_0,\hat{\bi{n}},\bi{k})&=&
\xi^1(\bi{k})e^{2i\phi}\int_0^{\tau_0} d\tau
S_P^{(T)}(k,\tau)\left(-\partial \mu - {2 \over 1-\mu^2}\right)^2 \left[
(1-\mu^2) (1+\mu)^2 e^{ix\mu}\right] \nonumber \\
	&=&\xi^1e^{2i\phi}(\bi{k})\int_0^{\tau_0} d\tau
S_P^{(T)}(k,\tau)[-{\hat{\cal E}}(x)+i{\hat{\cal B}}(x)]\left[
(1-\mu^2) e^{ix\mu}\right] \nonumber \\
\end{eqnarray}
where we introduced operators
${\hat{\cal E}}(x)=-12+x^2[1-\partial_x^2]-8x\partial_x $ and 
${\hat{\cal B}}(x)=8x+2x^2\partial_x$. Expressions for
the terms proportional to $\xi^2$ can be obtained analogously.

For tensor modes all three quantities $\Delta_T^{(T)}$,
$\Delta_{\tilde{E}}^{(T)}$ and 
$\Delta_{\tilde{B}}^{(T)}$ are non-vanishing and given by
\begin{eqnarray}
\Delta_T^{(T)}
(\tau_0,\hat{\bi{n}},\bi{k})&=&\Big[(1-\mu^2)e^{2i\phi}\xi^1(\bi{k})+
(1-\mu^2)e^{-2i\phi}\xi^2(\bi{k})\Big]
\int_0^{\tau_0} d\tau S_T^{(T)}(\tau,k)\; e^{ix\mu}
\nonumber \\
\Delta_{\tilde{E}}^{(T)}
(\tau_0,\hat{\bi{n}},\bi{k})&=&\Big[(1-\mu^2)e^{2i\phi}\xi^1(\bi{k})+
(1-\mu^2)e^{-2i\phi}\xi^2(\bi{k})\Big]{\hat{\cal E}}(x)
\int_0^{\tau_0} d\tau S_P^{(T)}(\tau,k)\; e^{ix\mu}
\nonumber \\
\Delta_{\tilde{B}}^{(T)}
(\tau_0,\hat{\bi{n}},\bi{k})&=&\Big[(1-\mu^2)e^{2i\phi}\xi^1(\bi{k})- 
(1-\mu^2)e^{-2i\phi}\xi^2(\bi{k})\Big]{\hat{\cal B}}(x)
\int_0^{\tau_0} d\tau S_P^{(T)}(\tau,k)\; e^{ix\mu}.
\label{tebT}
\end{eqnarray}
  From these expressions and equations (\ref{aeb}), (\ref{statxi})
one can explicitly show that
$B$ does not cross correlate with either $T$ or $E$.
 
The temperature power spectrum can be obtained easily in this
formulation,
\begin{eqnarray} 
C_{Tl}^{(T)}&=&
{4\pi \over 2l+1} \int k^2dk P_h(k)\sum_m \left|\int d\Omega 
Y^*_{lm}(\hat{\bi{n}})
\int_0^{\tau_0} d\tau
S_T^{(T)}(k,\tau)
\; (1-\mu^2) e^{2i\phi}e^{ix\mu}\right|^2
\nonumber \\
&=& 4\pi^2{(l-2)!\over (l+2)!}\int k^2dk P_h(k)\left|
\int_0^{\tau_0} d\tau S_T^{(T)}(k,\tau)
\int_{-1}^1 d\mu P^2_l(\mu)
\; (1-\mu^2) e^{ix\mu}\right|^2
\nonumber \\
&=& 4\pi^2{(l-2)!\over (l+2)!}\int k^2dk P_h(k)\left|
\int_0^{\tau_0} d\tau S_T^{(T)}(k,\tau)
\int_{-1}^1 d\mu {d^2 \over d\mu^2} P_l(\mu)
\; (1-\mu^2)^2 e^{ix\mu}\right|^2
\nonumber \\
&=& 4\pi^2{(l-2)!\over (l+2)!}\int k^2dk P_h(k)\left|
\int_0^{\tau_0} d\tau S_T^{(T)}(k,\tau)
\int_{-1}^1 d\mu {d^2 \over d\mu^2} P_l(\mu)
\; (1+\partial_x^2)^2 e^{ix\mu}\right|^2
\nonumber \\
&=& 4\pi^2{(l-2)!\over (l+2)!}\int k^2dk P_h(k)\left|
\int_0^{\tau_0} d\tau S_T^{(T)}(k,\tau)
\int_{-1}^1 d\mu P_l(\mu)
\; (1+\partial_x^2)^2 (x^2 e^{ix\mu})\right|^2
\nonumber \\
&=&(4\pi)^2{(l+2)!\over (l-2)!}\int k^2 dk P_h(k)\left|
\int_0^{\tau_0} d\tau 
S_T^{(T)}(k,\tau) {j_l(x)\over x^2}\right|^2,
\end{eqnarray}
where we used 
$Y_{lm}=[(2l+1)(l-m)!/(4\pi)(l+m)!]^{1/2}P_l^{m}(\mu)
e^{im\phi}$ and $P_l^m(\mu)=(-1)^m
(1-\mu^2)^{m/2}{d^m \over d\mu^m}P_l(\mu)$.
Note that the calculation involved in the last step is the
same as for the scalar polarization. The final expression agrees with
the expression given in \cite{sz}, which was obtained using 
the radial decomposition of the tensor eigenfunctions \cite{abbott}.
Although the final result is not new, 
the simplicity of the derivation presented here demonstrates the 
utility of this approach and will in fact be used to derive 
tensor polarization power spectra.

The expressions for the $E$ and $B$ power spectra are now easy to 
derive by noting that the angular dependence 
for $\Delta_{\tilde{E}}^{(T)}$ and 
$\Delta_{\tilde{B}}^{(T)}$ in (\ref{tebT}) are equal 
to those for $\Delta_T^{(T)}$.
The expressions only differ in 
the $\hat{\cal E}$ and $\hat{\cal B}$ operators 
that can be applied after the
angular integrals are done.
This way we obtain using equation (\ref{eblm})
\begin{eqnarray} 
C_{El}^{(T)}&=&
(4\pi)^2\int k^2dk P_h(k)\left|
\int_0^{\tau_0} d\tau 
S_P^{(T)}(k,\tau) {\hat{\cal E}}(x){j_l(x)\over x^2}\right|^2 \nonumber \\
&=&
(4\pi)^2\int k^2dkP_h(k)\left(
\int_0^{\tau_0} d\tau
S_P^{(T)}(k,\tau)\Big[-j_l(x)+j_l''(x)+{2j_l(x) \over x^2}
+{4j_l'(x) \over x}\Big]\right)^2 \nonumber \\
C_{Bl}^{(T)}&=&
(4\pi)^2\int k^2dk P_h(k)\left|
\int_0^{\tau_0} d\tau 
S_P^{(T)}(k,\tau) {\hat{\cal B}}(x){j_l(x)\over x^2}\right|^2 \nonumber \\
\nonumber \\
&=&
(4\pi)^2\int k^2dkP_h(k)\left(
\int_0^{\tau_0} d\tau
S_P^{(T)}(k,\tau)\Big[2j_l'(x)
+{4j_l \over x}\Big]\right)^2 
\end{eqnarray}

For computational purposes it is convenient to
further simplify these expressions by integrating by parts the 
derivatives $j_l'(x)$ and $j_l''(x)$. 
This finally leads to
\begin{eqnarray}
\Delta_{Tl}^{(T)}&=&\sqrt{(l+2)! \over (l-2)!}\int_0^{\tau_0} d\tau S_T^{(T)}(k,\tau){j_l(x) \over x^2} \nonumber \\
\Delta_{E,Bl}^{(T)}&=&\int_0^{\tau_0} d\tau S_{E,B}^{(T)}(k,\tau)j_l(x) \nonumber \\
S_E^{(T)}(k,\tau)&=&g\left(\Psi-{\ddot{\Psi}\over k^2}+{2\Psi \over x^2}
-{\dot{\Psi}\over kx}\right)-\dot{g}\left({2\dot{\Psi}\over k^2}+
{4 \Psi \over kx}\right)-2\ddot{g}{\Psi \over k^2}
 \nonumber \\
S_B^{(T)}(k,\tau)&=&g\left({4\Psi \over x}+{2\dot{\Psi}\over k}\right)+
2\dot{g} {\Psi \over k}. 
\label{et}
\end{eqnarray}
The power spectra are given by
\begin{eqnarray} 
C_{Xl}^{(T)}&=&
(4\pi)^2\int k^2dkP_h(k)\Big[\Delta^{(T)}_{Xl}(k)\Big]^2
\nonumber \\
C_{Cl}^{(T)}&=&
(4\pi)^2\int k^2dkP_h(k)\Delta^{(T)}_{Tl}(k)
\Delta^{(T)}_{El}(k),
\label{est}
\end{eqnarray}
where $X$ stands for $T$, $E$ or $B$.
Equations (\ref{et}) and (\ref{est}) are the main results of this section.

\section{Small scale limit}

In this section we derive the expressions for polarization 
in the small scale limit. The purpose of this section is to make
a connection with previous work on this subject 
\cite{crittenden93,prev,uros,kosowsky96} and to provide an estimate
on the validity of the small scale approximation.
In the small scale limit one considers only directions in the sky 
$\hat{\bi{n}}$ which are close to
$\hat{\bi{z}} $, in which case instead of spherical decomposition
one may use a plane wave expansion.
For temperature anisotropies
we replace 
\begin{equation}
\sum_{lm}a_{T,lm}Y_{lm}(\hat{\bi{n}}) \longrightarrow
\int d^2\bi{l} T(\bi{l})e^{i\bi{l}\cdot\bi{\theta}}, 
\end{equation}
so that
\begin{equation}
T(\hat{\bi{n}})=(2\pi)^{-2}\int d^2 \bi{l}\;\;
T(\bi{l})e^{i\bi{l} \cdot \bi{\theta}}. 
\end{equation}
To expand $s=\pm 2$ weighted
functions we use
\begin{eqnarray}
_2Y_{lm}=
 \left[{(l-2)!\over (l+2)!}\right]^{1\over 2}\edth^2 Y_{lm}
&\longrightarrow&(2\pi)^{-2}{1\over l^2}\edth^2 e^{i\bi{l} \cdot \bi{\theta}}
\nonumber \\ 
_{-2}Y_{lm}=
 \left[{(l-2)!\over (l+s)!}\right]^{1\over 2}(
\baredth^{2} Y_{lm}
&\longrightarrow&(2\pi)^{-2}{1\over l^2}\baredth^2 
e^{i\bi{l} \cdot \bi{\theta}},
\end{eqnarray}
which leads to the following expression
\begin{eqnarray}
(Q+iU)(\hat{\bi{n}})&=&-(2\pi)^2\int d^2 \bi{l}\;\;
[E(\bi{l})+iB(\bi{l})] {1\over l^2}\edth^2 
e^{i\bi{l} \cdot \bi{\theta}} \nonumber \\
(Q-iU)(\hat{\bi{n}})&=&-(2\pi)^2\int d^2 \bi{l}\;\;
[E(\bi{l})-iB(\bi{l})]
{1\over l^2}\baredth^2 
e^{i\bi{l} \cdot \bi{\theta}}.
\label{SSL1}
\end{eqnarray} 
    From equation (\ref{edth}) we obtain in the small scale limit
\begin{eqnarray}
{1\over l^2}\edth^2 
e^{i\bi{l} \cdot \bi{\theta}}&=& - e^{-2i(\phi-\phi_{l})}
e^{i\bi{l} \cdot \bi{\theta}} \nonumber \\
{1\over l^2}\baredth^2 
e^{i\bi{l} \cdot \bi{\theta}}&=& - e^{2i(\phi-\phi_{l})}
e^{i\bi{l} \cdot \bi{\theta}} \nonumber \\
\label{SSL2}
\end{eqnarray}
where $(l_x+il_y)=le^{i\phi_{l}}$.

The above expression was derived in the spherical basis where
$\hat{{\bi e}}_1=\hat{{\bi e}}_{\theta}$ and $\hat{{\bi e}}_2
=\hat{{\bi e}}_{\phi}$,
but in the small scale limit one can define a fixed basis in the sky
perpendicular to $\hat{\bi{z}}$,
$\hat{{\bi e}}_1'=\hat{{\bi e}}_{x}$ and $\hat{{\bi e}}_2'
=\hat{{\bi e}}_{y}$.
The Stokes parameters in the two coordinate
systems are related by
\begin{eqnarray}
(Q+iU)'&=&e^{-2i\phi}(Q+iU)\nonumber \\
(Q-iU)'&=&e^{2i\phi}(Q-iU).
\label{SSL3}
\end{eqnarray}
Combining equations (\ref{SSL1}-\ref{SSL3}) we find
\begin{eqnarray}
Q'(\bi{\theta})&=&(2\pi)^{-2}\int d^2 \bi{l}\;\;
[E(\bi{l}) \cos(2\phi_{l})
-B(\bi{l}) \sin(2\phi_{l})]
e^{i\bi{l} \cdot \bi{\theta}} 
\nonumber \\
U'(\bi{\theta})&=&(2\pi)^{-2}\int d^2 \bi{l}\;\;
[E(\bi{l}) \sin(2\phi_{l})
+B(\bi{l}) \cos(2\phi_{l})]
e^{i\bi{l} \cdot \bi{\theta}}. 
\label{QUreal}
\end{eqnarray}
These relations agree with those given in \cite{uros}, which were 
derived in the small scale approximation. As already shown there,
power spectra and correlation functions for $Q$ and $U$ used in 
previous work on this subject \cite{crittenden93,prev,kosowsky96}
can be simply derived from these
expressions. Of course,
for scalar modes $B^{(S)}(\bi{l})=0$, while for the tensor
modes both $E^{(T)}(\bi{l})$ and $B^{(T)}(\bi{l})$ combinations contribute.

\begin{figure}[t]
\vspace*{17.3 cm}
\voffset=200pt
%\centerline{\psfig{figure=fig1a_long.ps,height=6in}}
%\centerline{\psfig{figure=fig1b_long.ps,height=6in}}
%\centerline{\psfig{figure=fig1c_long.ps,height=6in}}
\caption{Comparison between exact calculation (solid lines) and 
small scale approximation (dashed lines) for standard CDM 
model with and without reionization. In the latter case we use optical
depth of 0.2. The reionized models are the upper curves on large scales.
The comparison is for scalar $E$ (a) and tensor $E$ (b) 
and $B$ (c) polarization power spectra. 
The spectra are in units of $T_0^2=(2.729K)^2$ and are normalized to COBE.
While the predictions agree
for large $l$ there are significant discrepancies in certain models
for small $l$.}
\includegraphics{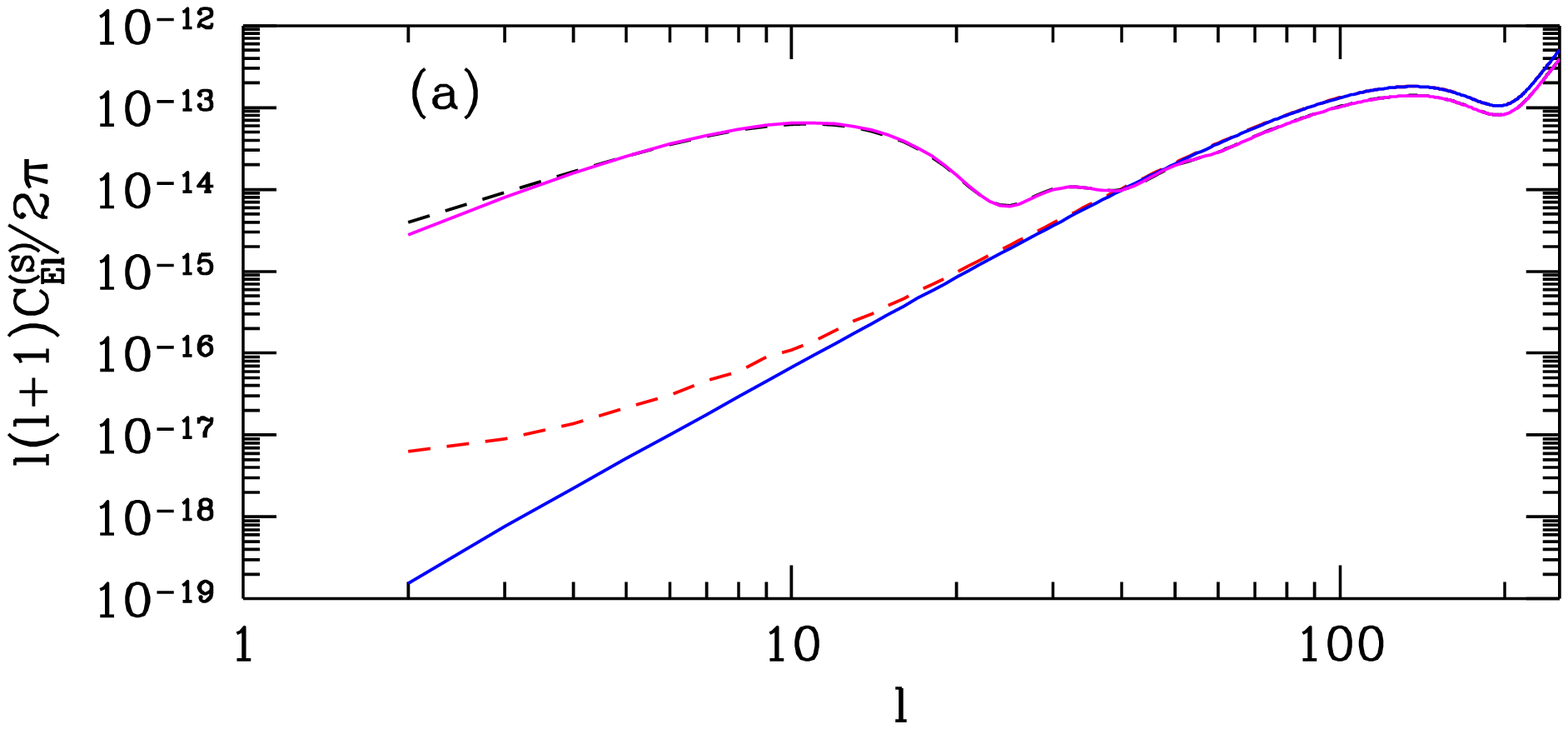}
\includegraphics{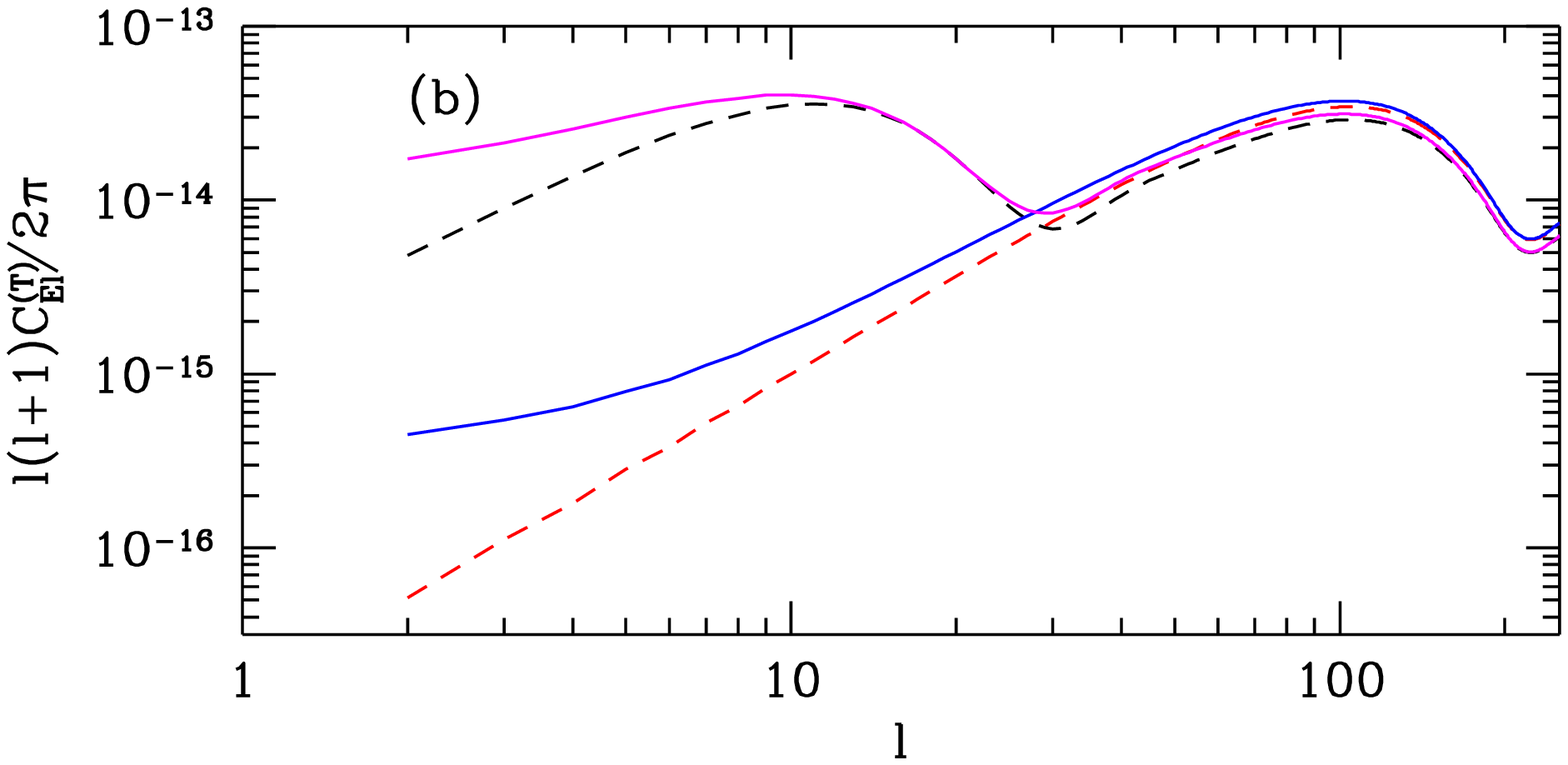}
\includegraphics{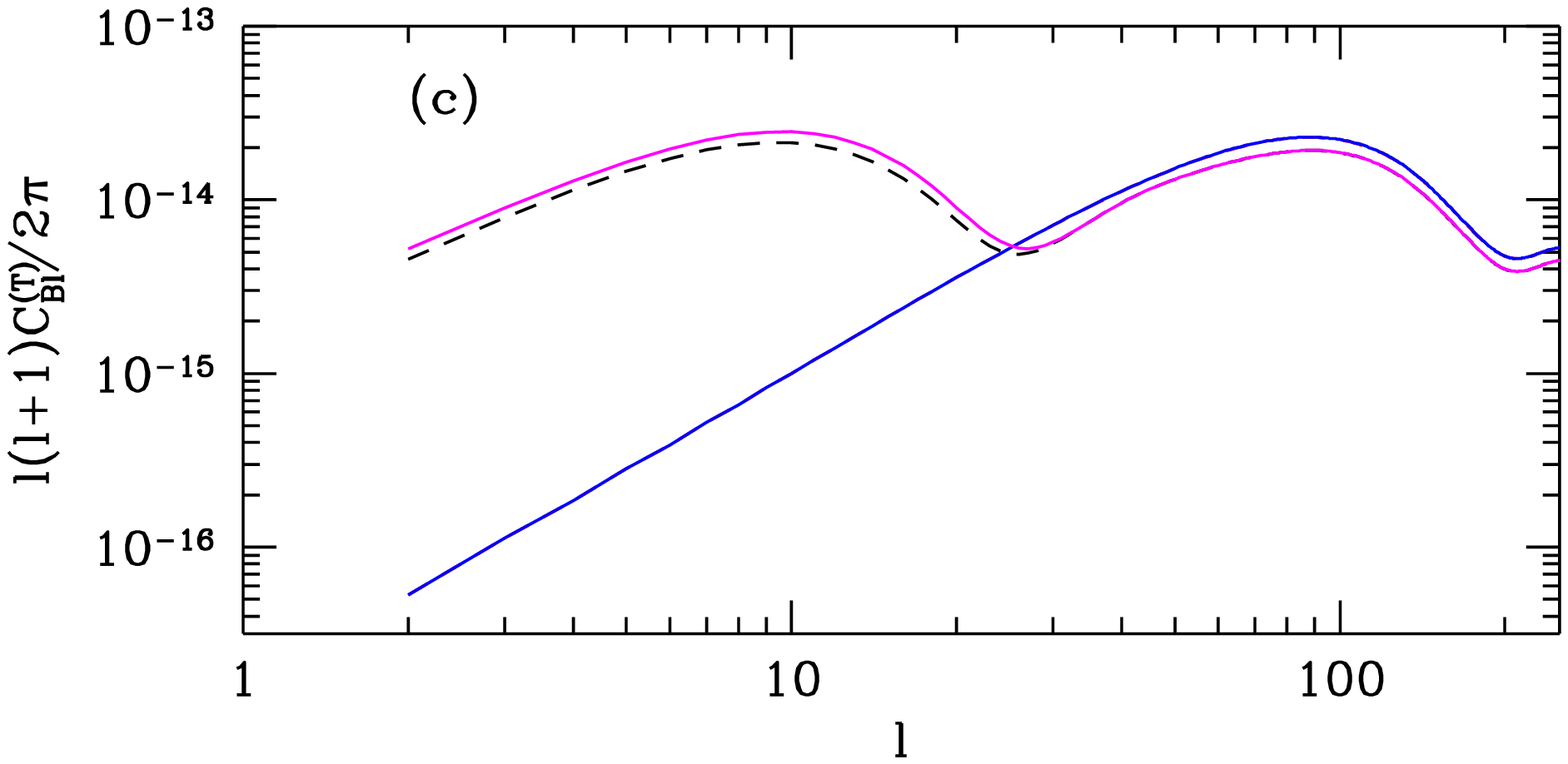}
\label{fig1}
\end{figure}

The expressions for $Q$ and $U$ (equation \ref{QUreal})
are easier to compute
in the small scale limit 
than the general expressions 
presented in this paper
(equation \ref{Pexpansion}), because
Fourier analysis allows one to use Fast Fourier Transform
techniques. In addition, the characteristic signature of scalar polarization
is simple to understand in this limit and can in principle be directly
observed with the interferometer measurements \cite{uros}. 
On the other hand, the exact power spectra derived in this
paper (equations \ref{es}, 
\ref{esc} and \ref{et}, \ref{est}) are as
simple or even simpler to compute with the integral approach 
than their small scale analogs.
Note that this need not be the case if one uses the standard approach
where Boltzmann equation is first expanded in a hierarchical system of
coupled differential equations \cite{bond87}.
In Fig. \ref{fig1} we compare the exact power spectrum (solid lines)
with the one derived in the small scale approximation (dashed lines), 
both for scalar $E$ (a) and tensor $E$ (b) and $B$ (c) combinations. 
The two models are standard CDM with and without reionization. The latter
boosts the amplitude of polarization on large scales.
The integral solution for scalar polarization in the small scale approximation
was given in \cite{sz} and is actually more complicated that the 
exact expression presented in this paper. 
In the reionized case the small scale approximation 
agrees well with exact calculation even at very large scales, while in 
the standard recombination scenario 
there are significant differences for $l<30$. 
Even though the relative error is large in this case, 
the overall amplitude
on these scales is probably too small to be observed. 

For tensors the small scale approximation
results in equation (\ref{et}) without the terms that contain $x^{-1}$ or
$x^{-2}$. Because $j_l(x) \sim 0$ for $x<l$ 
these terms are suppressed by $l^{-1}$ and 
$l^{-2}$, respectively, and are negligible compared to other terms 
for large $l$.
The small scale approximation agrees well with the exact 
calculation for $B$ combination (Fig. \ref{fig1}c), specially for 
the no-reionization model. For the $E$ combination the agreement is 
worse and there are notable discrepancies between the two even at 
$l \sim 100$. We conclude that
although the small scale expressions for the power spectrum 
can provide a good 
approximation in certain models, there
is no reason to use these instead of the exact expressions.
The exact 
integral solution for the power spectrum requires no additional 
computational expense 
compared to the small scale approximation and it should be used whenever 
accurate theoretical predictions are required. 
\section{Analysis of all-sky maps}

In this section we discuss issues related to simulating and
analyzing all-sky polarization and temperature
maps. This should be specially useful
for future satellite missions \cite{map,cobra}, which will measure 
temperature anisotropies and polarization over 
the whole sky
with a high angular resolution. 
Such an all-sky analysis will be of particular importance 
if reionization and tensor fluctuations
are important, in which case polarization will give
useful information on large angular scales, where Fourier analysis 
(i.e. division of
the sky into locally flat patches) is not possible. In addition, it is
important
to know how to simulate an all-sky map which preserves proper correlations
between neighboring patches of the sky and with which small scale
analysis can be tested for possible biases.

To make an all-sky map we need to generate the multipole moments
$a_{T,lm}$, $a_{E,lm}$ and $a_{B,lm}$.
This can be done by a generalization of the method given
in \cite{uros}. For each $l$ one diagonalizes
the correlation matrix $M_{11}=C_{Tl}$,
$M_{22}=C_{El}$, $M_{12}=M_{21}=C_{Cl}$ and generates 
from a normalized gaussian distribution two pairs of
random numbers (for real and imaginary components of $a_{l\pm m}$).
Each pair is multiplied with the square root of
eigenvalues of $M$ and rotated back to the original frame.
This gives a realization of $a_{T,l\pm m}$ and $a_{E,l\pm m}$ with correct
cross-correlation properties. For $a_{B,l\pm m}$ the procedure is simpler,
because it does not cross-correlate with either $T$ or $E$, so a pair
of gaussian random variables is multiplied with $C_{Bl}^{1/2}$ to
make a realization of $a_{B,l\pm m}$. Of course, for scalars $a_{B,lm}=0$.

Once $a_{E,lm}$ and $a_{B,lm}$ are generated we can form their linear
combinations $a_{2,lm}$ and $a_{-2,lm}$, which are equal in the scalar case.
Finally, to make a map of $Q(\hat{\bi n})$ and $U(\hat{\bi n})$ in the
sky we perform the sum in equation (\ref{Pexpansion}), using the explicit
form of
spin-weighted harmonics ${}_sY_{lm}(\hat{\bi n})$ 
(equation \ref{expl}).
To reconstruct the polarization
power spectrum from a map of $Q(\hat{\bi n})$ and
$U(\hat{\bi n})$ one first combines them in $Q+iU$ and $Q-iU$ to obtain
spin $\pm 2$ quantities. Performing the integral
over ${}_{\pm 2}Y_{lm}$ (equation \ref{alm})
projects out ${}_{\pm 2}a_{lm}$, from which $a_{E,lm}$ and $a_{B,lm}$
can be obtained.

Once we have the multipole moments we can 
construct various power spectrum estimators and analyze their
variances. In the case of full sky coverage one may generalize the
approach in \cite{knox95} to estimate the variance in the power spectrum 
estimator in the presence of noise. We will assume that we are given a
map of temperature
and polarization with $N_{pix}$ pixels and 
that the noise is uncorrelated from pixel to pixel
and also between $T$, $Q$ and $U$. 
The rms noise in the temperature is
$\sigma_T$ and that in $Q$ and $U$ is $\sigma_P$. If temperature and
polarization are obtained from the same experiment by adding and
subtracting the intensities between two orthogonal polarizations then
the rms noise in temperature and polarization
are related by $\sigma_T^2=\sigma_P^2/2$
\cite{uros}.

Under these conditions and using the orthogonality
of the $\;_sY_{lm}$ we obtain the statistical property of noise,
\begin{eqnarray}
\langle (a_{T,lm}^{{\rm noise}})^{*}a^{{\rm noise}}_{T,l^{\prime}m^{\prime}}\rangle 
&=& {4\pi \sigma_T^2 \over N_{pix}} 
\delta_{l l^{\prime}} \delta_{m m^{\prime}}
\nonumber \\
\langle (a^{{\rm noise}}_{2,lm})^{*}a^{{\rm noise}}_{2,l^{\prime}m^{\prime}}\rangle 
&=& {8\pi \sigma_P^2 \over N_{pix}} 
\delta_{l l^{\prime}} \delta_{m m^{\prime}}
\nonumber \\
\langle (a^{{\rm noise}}_{-2,lm})^{*}a^{{\rm noise}}_{-2,l^{\prime}m^{\prime}}\rangle 
&=& {8\pi \sigma_P^2 \over N_{pix}} 
\delta_{l l^{\prime}} \delta_{m m^{\prime}}
\nonumber \\
\langle (a^{{\rm noise}}_{-2,lm})^{*}a^{{\rm noise}}_{2,l^{\prime}m^{\prime}}\rangle 
&=& 0,
\end{eqnarray}
where by assumption there are no correlations between the noise in 
temperature and polarization. 
With these and equations
(\ref{aeb},\ref{stat}) we find 
\begin{eqnarray}
\langle a_{T,lm}^{*}a_{T,l^{\prime}m^{\prime}}\rangle 
&=& (C_{Tl} e^{-l^2 \sigma_b^2} + w_T^{-1})
\delta_{l l^{\prime}} \delta_{m m^{\prime}}
\nonumber \\
\langle a_{E,lm}^{*}a_{E,l^{\prime}m^{\prime}}\rangle 
&=& (C_{El} e^{-l^2 \sigma_b^2} + w_P^{-1}) 
\delta_{l l^{\prime}} \delta_{m m^{\prime}}
\nonumber \\
\langle a_{B,lm}^{*}a_{B,l^{\prime}m^{\prime}}\rangle 
&=& (C_{Bl} e^{-l^2 \sigma_b^2} +w_P^{-1}) 
\delta_{l l^{\prime}} \delta_{m m^{\prime}}
\nonumber \\
\langle a_{E,lm}^{*}a_{T,l^{\prime}m^{\prime}}\rangle 
&=& C_{Cl} e^{-l^2 \sigma_b^2} 
\delta_{l l^{\prime}} \delta_{m m^{\prime}}
\nonumber \\
\langle a_{B,l^\prime m^\prime}^{*} a_{E,lm}\rangle&=&
\langle a_{B,l^\prime m^\prime}^{*} a_{T,lm}\rangle=
0. 
\label{almvar}
\end{eqnarray}
For simplicity we characterized the beam smearing by 
$e^{l^2 \sigma_b /2}$ where $\sigma_b$ is the gaussian size of the beam
and we defined $w_{T,P}^{-1}=4\pi\sigma_{T,P}^2/N$ \cite{uros,knox95}.

The estimator for the temperature power spectrum is \cite{knox95},
\begin{eqnarray}
\hat{C}_{Tl}&=&\left[\sum_m{ |a_{T,lm}|^2 \over 2l+1} -  w^{-1}_T 
\right]e^{l^2\sigma_b^2}
\end{eqnarray}
Similarly for polarization and cross correlation the optimal
estimators are given by \cite{uros}
\begin{eqnarray}
\hat{C}_{El}&=&\left[\sum_m{ |a_{E,lm}|^2 \over 2l+1} -  w^{-1}_P 
\right]e^{l^2\sigma_b^2}\nonumber \\
\hat{C}_{Bl}&=&\left[\sum_m{ |a_{B,lm}|^2 \over 2l+1}-  w^{-1}_P 
\right]e^{l^2\sigma_b^2}\nonumber \\
\hat{C}_{Cl}&=&\left[\sum_m{ (a_{E,lm}^{*}a_{T,lm}+a_{E,lm}a_{T,lm}^{*}) 
\over 2(2l+1)}\right]e^{l^2 \sigma_b^2}.
\end{eqnarray}

The covariance matrix between the different estimators, 
${\rm Cov }(\hat{X}\hat{X}^{\prime})=\langle (\hat X - \langle \hat X \rangle)
(\hat X^{\prime} - \langle \hat X^{\prime} \rangle)\rangle$
is easily calculated using equation (\ref{almvar}). 
The diagonal terms are given by
\begin{eqnarray}
{\rm Cov }(\hat{C}_{Tl}^2)&=&{2\over 2l+1}(\hat{C}_{Tl}+
w_T^{-1}e^{l^2 \sigma_b^2})^2
\nonumber \\
{\rm Cov }(\hat{C}_{El}^2)&=&{2\over 2l+1}(\hat{C}_{El}+
w_P^{-1}e^{l^2 \sigma_b^2})^2
\nonumber \\
{\rm Cov }(\hat{C}_{Bl}^2)&=&{2\over 2l+1}(\hat{C}_{Bl}+
w_P^{-1}e^{l^2 \sigma_b^2})^2
\nonumber \\
{\rm Cov }(\hat{C}_{Cl}^2)&=&{1\over 2l+1}\left[\hat{C}_{Cl}^2+
(\hat{C}_{Tl}+w_T^{-1}e^{l^2 \sigma_b^2})
(\hat{C}_{El}+w_P^{-1}e^{l^2 \sigma_b^2})\right].
\end{eqnarray}
The non-zero off diagonal terms are
\begin{eqnarray}
{\rm Cov }(\hat{C}_{Tl}\hat{C}_{El})&=&{2\over 2l+1}\hat{C}_{Cl}^2
\nonumber \\
{\rm Cov }(\hat{C}_{Tl}\hat{C}_{Cl})&=&{2\over 2l+1}\hat{C}_{Cl}
(\hat{C}_{Tl}+w_T^{-1}e^{l^2 \sigma_b^2})
\nonumber \\
{\rm Cov }(\hat{C}_{El}\hat{C}_{Cl})&=&{2\over 2l+1}\hat{C}_{Cl}
(\hat{C}_{El}+w_P^{-1}e^{l^2 \sigma_b^2}).
\end{eqnarray}
These expressions agree in the small scale limit with those given in 
\cite{uros}. Note that the theoretical analysis is more
complicated if all four power spectrum estimators are used to deduce
the underlying cosmological model. For example, to test the sensitivity of 
the spectrum on the underlying parameter one uses the Fisher information
matrix approach \cite{jungman}. If only temperature information is
given then for each $l$ a derivative of the temperature
spectrum with respect to the parameter under investigation is computed
and this information is then summed over all $l$ weighted  
by ${\rm Cov }^{-1}(\hat{C}_{Tl}^2)$. 
In the more general case discussed here instead of a single derivative 
we have a vector of four derivatives and the
weighting is given by the inverse of the covariance matrix,
\begin{equation}
\alpha_{ij}=\sum_l \sum_{X,Y}{\partial \hat{C}_{Xl} \over \partial s_i}
{\rm Cov}^{-1}(\hat{C}_{Xl}\hat{C}_{Yl}){\partial \hat{C}_{Yl} \over \partial s_j},
\end{equation}
where $\alpha_{ij}$ is the Fisher information or curvature 
matrix, ${\rm Cov}^{-1}$ is the inverse of the covariance matrix,
$s_i$ are the cosmological parameters one would like to 
estimate and $X,Y$ stands for $T,E,B,C$. For each $l$ one has to
invert the covariance matrix and sum over $X$ and $Y$,
which makes the numerical evaluation of this expression somewhat
more involved.

\section{Conclusions}

In this paper 
we developed the formalism for an all-sky analysis of polarization 
using the theory of spin-weighted functions. 
We show that one can define rotationally invariant 
electric and magnetic-type parity fields $E$ and $B$ 
from the usual $Q$ and $U$ Stokes parameters.
A complete statistical 
characterization of CMB anisotropies 
requires four correlation functions,
the auto-correlations of $T$, $E$ and $B$ and the cross-correlation 
between $E$ and $T$. The pseudo-scalar nature of $B$
makes its cross-correlation with $T$ and $E$ vanish.  
For scalar modes $B$ field vanishes. 

Intuitive understanding of these results
can be obtained by considering polarization created by each plane 
wave given by direction $\bi{k}$. Photon propagation
can be described by scattering through a plane-parallel medium.
The cross-section only depends on the angle between photon 
direction $\bi{\hat{n}}$ and $\bi{k}$, so for a local coordinate system 
oriented in this direction only $Q$ Stokes parameter will be 
generated, while $U$ will vanish by symmetry arguments \cite{chandra}. 
In the real universe one has to consider a superposition of plane waves
so this property does not hold in real space. However, by performing
the analog of a plane wave expansion on the sphere this property becomes
valid again and leads to the vanishing of $B$ in the scalar case.
For tensor perturbations this is not true even in this 
$\bi{k}$ dependent frame, because each plane
wave consists of two different independent ``polarization'' states, which 
depend not only on the direction of plane wave, but also
on the azimuthal angle perpendicular to $\bi{k}$. The symmetry 
above is thus explicitly broken. Both $Q$ and $U$ are generated
in this frame and, equivalently, both $E$ and $B$ are generated in general.

Combining the formalism of spin-weighted functions and
the line of sight solution of the Boltzmann equation  
we obtained the exact expressions for the power spectra both 
for scalar and tensor modes. We present their numerical evaluations
for a representative set of models. A numerical implementation of the
solution is publicly available and can be obtained from the
authors \cite{cmbfast}.
We also compared the exact solutions to their analogs
in the small scale approximation obtained previously. While the latter 
are accurate for all but the largest angular scales, the simple form 
of the exact solution suggests that the small scale approximation 
should be replaced with the exact solution for all calculations.
If both scalars and tensors are contributing to a particular
combination then the power spectrum for that combination is obtained
by adding the individual contributions. Cross-correlation terms
between different types of perturbations vanish after the integration
over azimuthal 
angle $\phi$ both for the temperature and for the $E$ and $B$ polarization,
as can be seen from equations (\ref{tilEs}) and (\ref{tebT}). 
This result holds even for the defect models, where the same
source generates scalar, vector and tensor perturbations.

In summary, 
future CMB satellite missions will produce all-sky maps of polarization
and these
maps will have to be analyzed using techniques similar to the one 
presented in this paper. Polarization measurements have the sensitivity
to certain cosmological parameters which is not achievable from the 
temperature 
measurements alone. This sensitivity is particularly important on 
large angular scales, where previously used approximations break down
and have to be replaced with the exact expressions for the polarization
power spectra presented in this paper.

\acknowledgments
We would like to thank D. Spergel for helpful discussions.
U.S. acknowledges 
useful discussions with M. Kamionkowski, A.
Kosowsky and A. Stebbins.

\appendix
\section{Spin-weighted functions}

In this Appendix we review the theory of spin-weighted functions
and their expansion in spin-s spherical harmonics. This was used
in the main text to make an all-sky expansion of Stokes $Q$ 
and $U$ Stokes parameters. The main application of these functions
in the past was in the theory of gravitational wave radiation (see e.g. 
\cite{thorne}).
Our discussion follows closely that of Goldberg et al. \cite{goldberg67},
which is based on the work by Newman and Penrose \cite{np}.
We refer to these references for a more detailed discussion.

For any given direction on the sphere
specified by the angles $(\theta,\phi)$, one can define three
orthogonal vectors, one radial and two tangential to
the sphere. Let us denote the radial 
direction vector with ${\bi n}$
and the tangential with $\hat{{\bi e}}_1$,  $\hat{{\bi e}}_2$.
The latter two are only defined up to 
a rotation around ${\bi n}$.

A function $\;_sf(\theta,\phi)$
defined on the sphere is said to have spin-s if under
a right-handed rotation of ($\hat{{\bi e}}_1$,$\hat{{\bi e}}_2$)
by an  angle $\psi$  it transforms as 
$\;_s f^{\prime}(\theta,\phi)=e^{-is\psi}\;_sf(\theta,\phi)$.
For example, given an arbitrary vector ${\bf a}$ on the sphere the
quantities ${\bf a}\cdot \hat{{\bi e}}_1+i{\bf a}\cdot\hat{{\bi e}}_2$, 
${\bf n} \cdot {\bf a}$ and ${\bf a}\cdot 
\hat{{\bi e}}_1-i{\bf a}\cdot\hat{{\bi e}}_2$ have spin
1,0 and -1 respectively.  
Note that we use a different convention for
rotation than Goldberg et al.
\cite{goldberg67} to agree with the previous literature on polarization.

A scalar field on the sphere can be expanded in spherical 
harmonics, $Y_{lm}(\theta,\phi)$, which form  
a complete and orthonormal basis. These functions
are not appropriate to expand spin weighted functions with $s\neq 0$.
There exist analog sets of functions  
that can be used to expand spin-s functions, the so called spin-s 
spherical harmonics $\;_sY_{lm}(\theta,\phi)$. These sets of functions
(one set for each particular spin) satisfy the same completness and 
orthogonality relations,
\begin{eqnarray}
\int_0^{2\pi} d\phi \int_{-1}^1 d\cos \theta \;_sY_{l^{\prime}m^{\prime}}^*(\theta,\phi)
\;_sY_{lm}(\theta,\phi)&=&\delta_{l^{\prime}l}\delta_{m^{\prime}m} 
\nonumber \\
\sum_{lm} \;_sY_{lm}^*(\theta,\phi)
\;_sY_{lm}(\theta^{\prime},\phi^{\prime})
&=&\delta(\phi-\phi^{\prime})\delta(\cos\theta-\cos\theta^{\prime}).
\end{eqnarray}   

An important property of spin-s functions is that there exists
a spin raising (lowering) 
operator $\edth$ ($\baredth$) with the property of 
raising (lowering) the spin-weight of a function,
$(\edth \,_sf)^{\prime}=e^{-i(s+1)\psi}
\edth \,_sf$, $(\baredth \,_sf)^{\prime}=e^{-i(s-1)\psi}
\baredth \,_sf$. Their explicit expression is given by
\begin{eqnarray}
\edth \;_sf(\theta,\phi)&=&-\sin^{s}(\theta)
\left[{\partial \over \partial\theta}+i\csc(\theta)
{\partial \over \partial\phi} \right]\sin^{-s}(\theta)\;_sf(\theta,\phi) 
\nonumber \\
\baredth \;_sf(\theta,\phi)&=&-\sin^{-s}(\theta)
\left[{\partial \over \partial\theta}-i\csc(\theta)
{\partial \over \partial\phi} \right]\sin^{s}(\theta)\;_sf(\theta,\phi) 
\label{edth}
\end{eqnarray}

In this paper we are interested in polarization, which is a quantity 
of spin $\pm 2$. 
The $\baredth$ and $\edth$ 
operators acting twice on a function $\;_{\pm 2}f(\mu,\phi)$ 
that satisfies $\partial_{\phi}\;_sf=im\;_sf$ can be expressed as
\begin{eqnarray} 
\baredth^2 \;_2f(\mu,\phi)&=&
\left(-\partial \mu + {m \over 1-\mu^2}\right)^2 \left[
(1-\mu^2) \;_2f(\mu,\phi)\right]  \nonumber\\ 
\edth^2 \;_{-2}f(\mu,\phi)&=&
\left(-\partial \mu - {m \over 1-\mu^2}\right)^2 \left[(1-\mu^2) \;_{-2}
f(\mu,\phi)\right], 
\label{operators1}
\end{eqnarray}
where $\mu=\cos(\theta)$.
With the aid of these operators one can express  
$_sY_{lm}$ in terms of the spin zero spherical harmonics $Y_{lm}$,
which are the usual spherical harmonics,
\begin{eqnarray}
_sY_{lm}&=&
 \left[{(l-s)!\over (l+s)!}\right]^{1\over 2}\edth^s Y_{lm}
\;\;\; , (0 \leq s \leq l) 
\nonumber \\ 
_sY_{lm}&=&
 \left[{(l+s)!\over (l-s)!}\right]^{1\over 2}(-1)^s 
\baredth^{-s} Y_{lm}
\;\;\; ,(-l \leq s \leq 0). 
\end{eqnarray}

The following properties of spin-weighted harmonics are also useful
\begin{eqnarray}
\:_sY^*_{lm}&=&(-1)^{s}{}_{-s}Y_{l-m} \nonumber \\
\edth \:_sY_{lm}&=&\left[(l-s)(l+s+1)\right]^{1\over 2}
\:_{s+1}Y_{lm} \nonumber \\
\baredth \:_sY_{lm}&=&-\left[(l+s)(l-s+1)\right]^{1\over 2}
\:_{s-1}Y_{lm} \nonumber \\
\baredth\edth \:_sY_{lm}&=&-(l-s)(l+s+1)
\:_{s}Y_{lm} 
\label{propYs}
\end{eqnarray}
Finally, to construct a map of polarization one needs an explicit 
expression for the spin weighted functions,
\begin{eqnarray}
&{}&_sY_{lm}(\hat{n})=e^{im\phi}\Big[{(l+m)!(l-m)!\over (l+s)!(l-s)!}
{2l+1\over 4\pi}\Big]^{1/2}\sin^{2l}(\theta/2) \nonumber \\
&\times& \sum_r {l-s \choose r}{l+s \choose r+s-m}
(-1)^{l-r-s+m}{\rm cot}^{2r+s-m}(\theta/2).
\label{expl}
\end{eqnarray}

\end{document}